# New challenges for the pressure evolution of the glass temperature.

Sylwester J. Rzoska

Institute of High Pressure Physics Polish Academy of Sciences,
ul. Sokołowska 29/37, 01-142 Warsaw, Poland

**Correspondence:**
e-mail: sylwester.rzoska@unipress.waw.pl

**Abstract:**

The ways of portrayal of the pressure evolution of the glass temperature ($T_g$) beyond the dominated Simon-Glatzel-like pattern are discussed. This includes the possible common description of $T_g(P)$ dependences in systems described by $dT_g/dP > 0$ and $dT_g/dP < 0$. The latter is associated with the maximum of $T_g(P)$ curve hidden in the negative pressures domain. The issue of volume and density changes along the vitrification curve is also noted. Finally, the universal pattern of vitrification associated with the crossover from the low density (isotropic stretching) to the high density (isotropic compression) systems is proposed. Hypothetically, it may obey any glass former, from molecular liquids to colloids.

**Key Words:**
**Glass transition, high pressures, negative pressures, melting, universality, dynamics, glass forming ability.**

## 1. Introduction

Liquids on cooling solidify in the ordered crystalline state when passing the melting temperature ($T_m$). However, the fluidity can be also preserved below melting, down to the glass temperature $T_g \ll T_m$, where the solidification from the metastable ultraviscous/ultraslowing liquid to the solid amorphous glass state occurs (Berthier and Ediger, 2016; Rzoska et al., 2010; Donth, 2000). There are also numerous semi-crystalline systems where the vitrification is related to the solidification of one or few elements of symmetry: as examples can serve orientationally disordered crystals (ODICs, plastic crystals) (Drozd-Rzoska et al., 2006a and 2006b) or liquid crystals (Drozd-Rzoska, 2006; Drozd-Rzoska, 2009). For many systems passing $T_m$ without crystallization is associated with extreme temperature quench (Donth, 2000). However, there are also numerous glass formers where entering the metastable ultraviscous/ultraslowing domain is possible at any practical experimental cooling rate (Berthier and Ediger, 2016; Rzoska et al., 2010; Donth, 2000). Turnbull (Turnbull, 1969; Angell, 2008) formulated the broadly used empirical Glass Forming Ability (GFA) rule distinguishing poor ($T_g/T_m < 2/3$) and good glass formers ($T_g/T_m > 2/3$), linking $T_g$ and $T_m$. Notwithstanding, there is a notable difference between melting and vitrification: melting is related to the 'sudden and almost non-signaled' fusion on cooling whereas the glass transition is hallmarked by far previtreous super-Arrhenius (SA) changes of viscosity $\eta(T)$, primary relaxation time $\tau(T)$ or other dynamic properties (Berthier and Ediger, 2016; Rzoska et al., 2010; Donth, 2000). This opens the possibility of estimating the glass temperature from the analysis of previtreous effects well above $T_g$: as the general reference values $\eta(T_g) = 10^{13} Poise$ or $\tau(T_g) = 100s$ are assumed, since they correlate with the thermodynamic estimation (heat capacity or density scan) of $T_g$ related to $10 K/\min$ cooling rate (Rzoska et al., 2010; Donth, 2000). Although the ultimate form of portrayal $\tau(T, P)$ or $\eta(T, P)$ changes in previtreous ultraviscous/ultraslowing liquids near $T_g$ remains puzzling (Martinez-Garcia, 2013; Martinez-Garcia, 2014), most often the Vogel-Fulcher-Tammann (VFT) relation is used (Berthier and Ediger, 2016; Rzoska et al., 2010; Donth, 2000; Martinez-Garcia, 2013):

$$\tau(T) = \tau_0 \exp\left(\frac{D_T T_0}{T - T_0}\right), \qquad P = const \qquad (1)$$

where $\tau_0 = 10^{-14 \pm 2}$ is the prefactor, $T_0 < T_g$ is the VFT singular temperature and $D_T$ denotes the fragility strength coefficient linked to fragility metric $m = \left[d \log_{10} \tau / d(T_g/T)\right]_{T \to T_g}$ via dependence $D_T = 590/(m + \log_{10} \tau_0 / \log_{10}(T_g))$ (Böhmer, 1993).

The pressure counterpart of the VFT equation was first proposed for the analysis of viscosity changes in glycerol by Johari and Whalley (1972) and later for the primary relaxation time in dibutyl phthalate (Paluch et al., 1996):

$$\eta = \eta_0^P \exp\left(\frac{A}{P_0 - P}\right) \qquad \text{and} \qquad \tau = \tau_p^P \exp\left(\frac{A}{P_0 - P}\right) \qquad (2)$$

where: $T = const$, $\eta_0^P$ and $\tau_o^P$ denote prefactors, the amplitude $A = const$ and $P_0 > P_g$ is the "VFT-like" singular pressure.

However, eqs. (2) can reliably portray experimental data only for 'strong' (weakly non-Arrhenius) glass formers, assuming that measurements terminates at $P_{max} \ll P_0$. In ref. (Paluch, Rzoska et al., 1998) the relation able to portray previtreous 'dynamic effects' for arbitrary glass formers and ranges of pressures was proposed:

$$\tau(P) = \tau_0^P \exp\left(\frac{A(P)}{P_0 - P}\right) = \tau_0^P \exp\left(\frac{D_P P}{P_0 - P}\right) \tag{3}$$

In this relation the amplitude is pressure dependent $A = A(P) = D_P P$, and the fragility strength coefficient $D_P$ was introduced. It is notable that for the basic VFT eq. (1) the prefactor is 'approximately universal", i.e. $\tau_0 \approx 10^{-14\pm2} s$, whereas for eqs. (2) and (3) it ranges between $\tau_0^P \approx 10s$ and $\tau_0^P \approx 10^{-14} s$ (Drozd-Rzoska and Rzoska, 2006; Drozd-Rzoska et al. 2008). Such enormous discrepancy results from the location of the isotherm selected for tests. This can be illustrated via the 'general' Super-Arrhenius equation:

$$\tau(T,P) = \tau_0^P \exp\left(\frac{PV_a(P)}{RT}\right) = \tau_0 \exp\left(\frac{E_a(T)}{RT}\right)\exp\left(\frac{PV_a(P)}{RT}\right) = \tau_0 \exp\left(\frac{E_a(T) + PV_a(P)}{RT}\right) \tag{4}$$

The comparison of eqs. (3) and (4) yields $E_a(T) = RD_T / (1/T_0 - 1/T)$ and $V_a(P) = TD_P R / (P_0 - P)$ for VFT estimations of the activation energy and activation volume, respectively. Notwithstanding, the general forms of $E_a(T)$ and $V_a(P)$ are not known. The solution of the problem of the poorly defined prefactor $\tau_0^P$ in eqs. (2) and (3) was proposed in refs. (Drozd-Rzoska and Rzoska, 2006; Drozd-Rzoska et al. 2008) by introducing the equation:

$$\tau(P) = \tau_0 \exp\left(\frac{D_P'(P - P_{Sp})}{P_0 - P}\right) = \tau_0 \exp\left(\frac{D_P' \Delta P}{P_0 - P}\right) \tag{5}$$

This dependence takes into account that the liquid state terminates at the absolute stability limit pressure (spinodal $P_{Sp}$), in negative pressures domain. The ultimate description needs both positive (isotropic compression, hydrostatic pressures, $P > 0$) and negative pressures (isotropic stretching, $P < 0$) domains (Imre et al, 2002). For eq. (5) the prefactor $\tau_0 = \tau(P_{Sp}) \approx 10^{-12} s$, for arbitrary isotherm. When comparing eqs. (3) and (5) worth noting is that the latter can penetrate negative pressures domain but fragility strength coefficients are different: $D_P / D_P' = P_0 / (P_0 - P_{Sp})$ (Drozd-Rzoska and Rzoska, 2006; Drozd-Rzoska et al. 2008).

The characterization of $T_g(P)$ dependence has a notable impact on the behavior under atmospheric pressure, being included via the coefficient $dT_g(P)/dP$ in numerous relations (Rzoska et al., 2010; Donth, 2000; Floudas et al., 2015; Rzoska and Mazur, 2007). The reliable knowledge of $T_g(P)$ description seems to be essential for silicate glasses, in which practically important features are created due to the high pressure − high temperature annealing with induced 'exotic' features preserved after decompressing. They are, for instance: (i) the notable increase of density, (ii) hardness and (iii) anty-cracking ability (Smedskjaer et al., 2014; Januchta et al., 2016; Svenson et al., 2017). Still puzzling is the description of $T_g(P)$ behavior in systems where $dT_g/dP < 0$ (Donth, 2000; Drozd-Rzoska, Rzoska and Imre, 2007; Drozd-Rzoska, Rzoska and Roland, 2007). All these show that the reliable and effective portrayal of the pressure evolution of the glass temperature can constitute one of milestones in dealing with the glass transition. This report presents the resume of this issue, supplemented by possible extensions beyond the current state-of-the art.

## 2. Parameterization of the pressure evolution of melting and glass temperatures

There are several relations for describing the pressure evolution of melting temperature: the most popular is the Simon-Glatzel (SG) equation, due its simple form and the limited number of fitted parameters (Simon and Glatzel, 1929; Skripov and Faizulin, 2006):

$$T_m(P) = T_0\left(1 + \frac{P}{a}\right)^{1/b} \qquad (6)$$

where $T_0$, $a$ and $b$ are adjustable parameters.

It can be derived from the Clausius-Clapeyron (C-C) equation $dT/dP = T\Delta V/\Delta H = \Delta V/\Delta S$, where $\Delta V$, $\Delta H$ and $\Delta S$ are for the volume, enthalpy and entropy changes at the transition, assuming $(dT/dP)_{fusion} = a + bP$ (Skripov and Faizulin, 2006). This relation describes melting, where the 'sudden and sharp' change of volume or density ($\Delta V$, $\Delta \rho$) and the heat capacity takes place. However, the C-C equation can be linked to any fusion phenomenon, provided it is associated with detectable changes in heat capacity and volume/density. This occurs also for the glass transition temperature, although the transformation is 'stretched' in temperature or pressure and occurs between the disordered (ultravisous) liquid and the disordered solid (glass), as exemplified in Fig. 1.

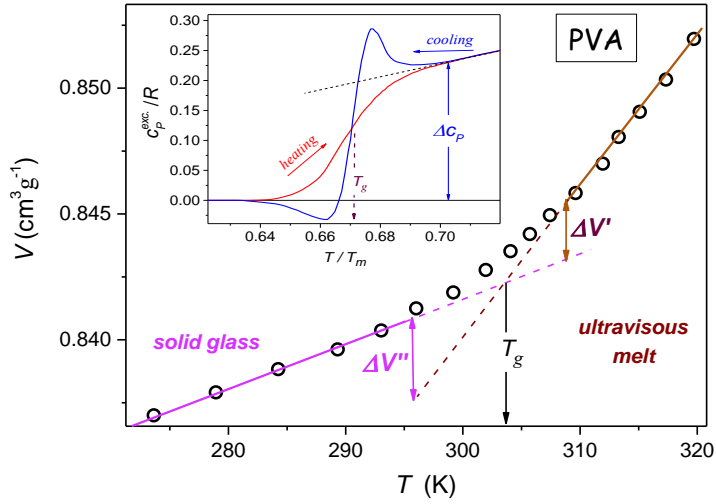

**Fig. 1** The temperature dependence ($P = 0.1$ MPa) of the proper volume $V = 1/\rho$, $\rho$ denotes density, for polyvinyl acetate (PVA) in the ultraviscous and solid amorphous phases. Dashed lines show extrapolations of the experimental behavior remote from the 'stretched' glass transition domain $T_g$. The apparent discontinuity of the volume can be estimated as $\Delta V' = 0.0021 cm^3 g^{-1}$ and $\Delta V'' = 0.0030 cm^3 g^{-1}$ (double arrows in the plot). The inset, based on data from ref. (McKinney, 1974; Tropin, 2012; Roland and Casalini, 2003), is for the excess of the specific heat $\Delta c_p(T) = c_P^{melt}(T) - c_p^{solid}(T)$, over the behavior in the solid stated remote from $T_g$ $c_p^{solid}(T) = a + bT$ described the behavior well below. The resulting discontinuity $\Delta c_p(T)/R = 0.23$. Data in Fig. 1 are for 10 K/min. cooling / heating rate.

As mentioned above the 'reasonable' metric of the glass transition is the isochronal or isoviscous condition $\tau(T_g, P_g) = 100s$ or $\eta(T_g, P_g) = 10^{13}$ Poise (Donth, 2000). Generally, such condition is absent along the melting curve within the P-T plane (Skripov and Faizulin, 2006). However, the isochronal condition for $T_m(P)$ is clearly fulfilled if melting is associated with only one element of symmetry, as for the isotropic − nematic transition in liquid crystals (Roland, Bogoslovov et al., 2008). Heuristic similarities between melting and vitrification can be strengthen recalling the empirical link between $T_g$ and $T_m$, used as the indicator of the Glass Forming Ability (GFA): $T_g/T_m > 2/3$ (near-spherical molecules) and $T_g/T_m > 1/2$ (elongated molecules) (Donth, 2000; Turnbull, 1969; Angell, 2008). Consequently, one can expect that the pressure dependence of $T_m$ are paralleled by $T_g(P)$ evolution. Regarding the vitrification, S. Peter Andersson and Ove Andersson (AA) introduced the SG-type relation for describing the pressure evolution of the glass temperature in poly(propylene) glycol (Andersson and Andersson, 1998):

$$T_g(P) = k_1 \left(1 + \frac{k_2}{k_3} P\right)^{1/k_2} \tag{7}$$

where $k_1$, $k_2$ and $k_3$ are empirical, adjustable parameters.

The AA equation has become the key tool for describing $T_g(P)$ experimental data till nowadays (Drozd-Rzoska et al., 2007; Floudas et al., 2011; Roland et al., 2005; Rzoska et al., 2007 and 2010). This success was notably strengthen by its derivation within the Avramov-Milchev (AM) phenomenological model for vitrification (Avramov and Milchev, 1988; Roland and Casalini, 2003):

$$T_g(P) = \varepsilon T_0 \left(1 + \frac{P}{\Pi}\right)^{\beta/\alpha} \tag{8}$$

where the coefficient $\varepsilon = \left[30 \log_{10}(e) / (\log_{10}(\tau(T_g)) - \log_{10} \tau_0)\right]^{1/\alpha}$

Notwithstanding, there is a discrepancy between eqs. (7) and (8) because the coefficient $\varepsilon \gg 1$. Worth recalling is also the criticism regarding the basic AM model dependence (Martinez-Garcia et al., 2013, Martinez-Garcia et al., 2014) $\tau(T) = \tau_0 \exp(A/T^D)$ or $\eta(T) = \eta_0 \exp(A/T^D)$, for $P = const$. For SG eq. (6) and AA eq. (7) always $dT_{g,m}/dP > 0$, i.e. $T_m(P)$ and $T_g(P)$ permanently increase with rising pressure. However, there are also systems $dT_{g,m}/dP < 0$. So far, their evidence for glass formers is still very limited: some of them are collected in Table I.

**Table I**  Examples of systems in which the application of pressure decreases the glass temperature ($dT_g/dP < 0$) [31-38]. For the dominant group of glass formers (molecular liquids, polymers, ..): $dT_g/dP > 0$ (Donth, 2000; Floudas et al. 2011, Roland et al., 2005).

| Glass Former | $dT_g/dP$, ($K/GPa$) | References |
|---|---|---|
| $CH_3COOLi + 10H_2O$ (ionic system) | -8.5 | (Kanno et al. 1981) |
| $LiOAc + 10xH_2O$ (ionic system) | -5 | (E. Williams, et al. 1977) |
| Water (model estimation) | -52 | (N. Giovambattista et al., 2012) |
| Albite (geo system) | -8.4 | (Bagdasarov et al., 2004) |
| Haplogranite (HPG8, geo system) | -45 | (Bagdasarov et al., 2004_ |
| Silicon (semiconductor) | -57 | (Deb et al. 2001) |
| $As_2Te_3$ (semiconductor) | -30 | (Ramesh, 2014) |
| $Ge_{20}Te_{80}$ (semiconductor) | -78 | (K. Ramesh et. al 2016). |
| RADP crystal (rubidium ammonium dihydrogen phosphate: paraelectric phase – glass state) | -41.5 | (Trybuła and Stankowski, 1998) |

It seems that such behavior may occur only for some strongly-bonded systems. Taking into account the clear evidence of systems with $T_m(P)$ maximum (Kechin, 1995; Kechin, 2001; Tonkov and Ponyatovsky, 2004), the similar behavior can be expected for $T_g(P)$. It is notable, that already a century ago (Tammann; 1903) it was indicated that the reversal melting $dT_m/dP > 0 \rightarrow (T_m^{max}, P_m^{max}) \rightarrow dT_m/dP < 0$ can be the general phenomenon.

The description of the reversal melting was first proposed by Rein and Demus (RD) (Rein and Demus, 1993; Demus and Pelzl, 1988) and subsequently by Kechin (K) (Kechin, 1995; Kechin 2001):

$$T_m(P) = T_0 \left(1 + \frac{P}{a}\right)^{1/b} \exp(-a_1 P) = R(P) \times D(P) \qquad (9)$$

where $a$, $b$ and $a_1$ are adjustable parameters. $R(P)$ denotes the SG-type 'rising' term and $D(P)$ is for the 'damping term'.

In subsequent decades eq. (9), recalled in references as the 'Kechin equation', became the key tool for describing experimental data associated with melting curve maximum (Rzoska at al., 2007 and 2010; Drozd-Rzoska, 2005; Skripov and Faizulin, 2006; Drozd-Rzoska, Rzoska and Imre, 2007). Regarding the meaning of parameters in eqs. (6-9) one can generalize the reasoning of Burakovsky et al. (Burakovsky et al., 2000; Burakovsky et al., 2003), who considered the volume-related compression factor:
$\eta = \Delta V_0 / \Delta V = (V(\pi) - V(P_0))/(V(P) - V(P_0))$ and linked it to the bulk (compressibility) modulus via $B = -\Delta V(d(\Delta P)/d(\Delta V)) = \eta d(\Delta P)/d\eta$, with the pressure dependence given as $B(P) = B_0 + B_0' P + ...$ and $\Delta P = P - P_0$:

$$\eta(P) = \left(1 + \frac{B_0'}{B_0} P\right)^{1/B_0'} \rightarrow T_m(P) = T_0 * (\eta)^{-1} = T_0 \left(1 + \frac{P}{B_0/B_0'}\right)^{1/B_0'} \qquad (10)$$

where the index '0' is related to the reference point ($T_0, P_0$).

Hence, taking as the reference the atmospheric pressure as the reference one can indicate the following meaning of parameters in eqs. (6) – (9) $a = B_0/B_0' = \pi$ and the power exponent $b = B_0'$. For SG and AA eqs. (6) and (7), as well as K&RD eq. (8), the reference has to be taken as $T_0 = T_{g,m}(P_0 = 0) \approx T_{g,m}(P_0 = 0.1 MPa)$. Other selections of $T_0$ yields non-optimal and effective values of fitted coefficients. In ref. (Skripov and Faizulin, 2006) as the general reference the triple point was proposed: and the $T_0 = T_{triple}$ and $P \rightarrow \Delta P = P - P_{triple}$ in the SG eq. (6). Notwithstanding, for many significant systems ($T_{triple}, P_{triple}$), Such general reference cannot be implemented for the glass transition. Drozd-Rzoska (Drozd-Rzoska, 2005; Drozd-Rzoska et al. 2007; Drozd-Rzoska et al. 2008) proposed as the reference arbitrary values ($T_0, P_0$) along melting or vitrification curves, taking $\Delta P = P - P_0$. Subsequently, assuming for the Clausius-Clapeyron equation along the melting or vitrification curve $(\Delta H/\Delta V)_{T_{g,m}, P_{g,m}} = (b\Pi + b\Delta P)/(1 - c(b\Pi + b\Delta P))$ the following relation was derived (Drozd-Rzoska, 2005):

$$T_{g,m}(P) = T_0 \left(1 + \frac{P - P_{Sp}}{\pi + P_0}\right) \times \exp\left(-\frac{P - P_0}{c}\right) = T_0 \left(1 + \frac{\Delta P}{\Pi}\right)^{1/b} \times \exp\left(-\frac{\Delta P}{c}\right) \qquad (11)$$

where $\Delta P = P - P_0$, $-\pi$ is the extrapolated, negative pressure for which $T_{g,m}(P \to -\pi) \to 0$: it correlates with the onset of $T_{Sp}(P_{Sp})$ absolute stability limit curve in negative pressures domain; $c$ is the damping pressure coefficient.

For small or moderate pressures one obtains the SG or AA type equation (Drozd-Rzoska, 2005; Drozd-Rzoska et al. 2007; Drozd-Rzoska et al. 2008):
'

$$T_{g,m}(P) \approx T_0 \left(1 + \frac{P - P_0}{\pi + P_0}\right) = T_0 \left(1 + \frac{\Delta P}{\Pi}\right)^{1/b} \qquad (12)$$

Eq. (11) is able to portray systems with the maximum of melting or vitrification curve, even if they are hidden in the negative pressures domain. It can be also applied for systems were $dT_{g,m}/dP < 0$. Eq. (12) can describe experimental data if $dT_{g,m}(P)/dP > 0$ and the set of data is well below the maximum of $T_{g,m}(P)$ curve. Both relations can be implemented in the negative pressures domain. Applying findings of Burakovsky et al. (Burakovsky et al., 2000) one obtains: $b = B_0'$ and $B_0/B_0' = P_0 + \pi$ and then $B_0 = B_0' P_0 + B_0' \pi$. The latter equation is in agreement with the empirical relation for the pressure evolution of the bulk modulus recalled above (Murnaghan, 1944).

There are few other approaches which starting from the C-C or related Lindemann relation (Skripov and Faizulin, 2006), developed for melting. They are briefly presented below, with indications of their applicability for the glass formation. All these is supplemented by few new formulas, resulted from such reasoning. Schlosser et al. (Schlosser et al., 1989) starting from the Lindemann relation $T_m = CV^{2/3}\Theta_D$ ($C$ is a constant, $\Theta_D$ is the Debye reduced temperature) (Lindemann, 1910; Skripove and Faizuli, 2006) and the definition of the Grüneisen parameter as $\gamma = (\partial \Theta_D/\partial V)_T = -\partial \ln \Theta_D/\partial \ln V$ (Grüneisen, 1913) obtained the relation focusing on the volume dependence of the melting temperature. Generalizing this dependence for the arbitrary fusion process one obtains:

$$T_{g,m}(V) = T_0 \left(\frac{V}{V_0}\right)^{2/3} \exp\left(2\gamma_0 \frac{V - V_0}{V_0}\right) = T_0 X^2 \exp\left(2\gamma_0 \frac{\Delta V}{V_0}\right) \qquad (13)$$

where the index '0' is for the zero-pressure (~atmospheric pressure) reference. Assuming for the $X^2 \approx 1 - 2\Delta V/3V_0 \approx \exp(-2\Delta V/3V_0)$ the following relation was derived (originally for melting):

$$T_{g,m}(V) = T_0 \exp\left(\frac{-2\Delta V}{3V_0}\right) \exp\left(2\gamma_0 \frac{\Delta V}{V_0}\right) \qquad (14)$$

One may expect that it is able to portray systems described both by $dT_{g,m}/dP > 0$ and $dT_{g,m}/dP < 0$. For small/moderate pressures eq. (14) can be reduced to the Kraut-Kennedy relation (Schlosser et al., 1989, Kraut and Kennedy, 1966), originally developed for melting:

$$T_{g,m} \approx T_0 \left[1 + 2(B_0 - 1/3)\frac{\Delta V}{V_0}\right] = T_0 (1 + C \Delta V/V_0) \qquad (15)$$

It can be converted to the density related dependence along melting or vitrification curves:

$$T_{g,m} \approx T_0 \left(1 + C \frac{\rho_0 - \rho}{\rho}\right) = T_0 \left(1 + C \frac{\Delta \rho}{\rho}\right) \qquad (16)$$

Linking eqs. (12) and (15) one obtains the relation for pressure induced volume changes along melting or vitrification curve:

$$\left(\frac{\Delta V}{V_0}\right)_{g,m} = \frac{(1+\Delta P/\Pi)^{1/b}-1}{C} \qquad (17)$$

This relation is in fair agreement with the Murnaghan equation, broadly used is earth sciences (Murnaghan, 1944; Skripov and Faizulin, 2006). Recalling the dependence $\Delta V/V_0 = \ln(1+\beta P)/\alpha$, where $\alpha = B'+1$ and $\beta = \alpha/B = (B_0'+1)/B$ eq. (15) can be converted to the SG- or AA- type equation (Schlosser et al, 1989):

$$T_{g,m}(P) \approx T_0(1+\beta P)^{2(B-1/3)/\alpha} \qquad (18)$$

It this relation the SG exponent $b = (B_0'+1)/(2(B_0-1/3))$, i.e. it differs from Burakovsky [Burakovsky et al., 2003) predictions.

Kumari and Dass (Kumari and Dass, 1988; Dass, 1995) also applied the framework of the Lindemann criterion (Lindemann, 1910) and workout the relation originally focused on the pressure evolution of the melting temperature, focusing on alkali metals:

$$\ln\left(\frac{T_{m,g}}{T_0}\right) = -2\alpha P + \left[2\left(C+\frac{\alpha}{\beta}\right)\ln(1+\beta P)\right] \qquad (19)$$

where $\alpha = (\gamma'/B')_{P_0,T_0}$, $\beta = (B'/B)_{P_0,T_0}$, $C = [(\gamma-1/3)/B']_{P_0,T_0}$, $\gamma$, $\gamma'$ and $B$, $B'$ stands for the Grüneisen parameter, bulk modulus and their first derivatives.

This relation can describe systems notably diverging from the SG pattern, including the cross over $dT_{g,m}/dP > 0 \to dT_{g,m}/dP < 0$. It can be also converted to the form coincided with Rein&Demus and Kechin eq. (8):

$$T_{m,g} = T_0(1+\beta P)^{2C+2\alpha/\beta} \exp(-2\alpha P) \qquad (20)$$

The coefficient $\alpha = \gamma'/B'$, what makes it possible to define the 'damping pressure' parameter in DR eq. (11): $c = B'/2\gamma'$. Eq. (20) can be reduced to the SG or AA forms assuming $\alpha = 0$ (Dass, 1995), i.e. $\gamma(P) = const$ in the given range of pressures:

$$T_{m,g}(P) = T_0(1+\beta P)^{2C} \qquad (21)$$

It is also notable that eq. (19) makes it possible to estimate the location of the maximum of $T_{g,m}(P)$ curves as $P_{g,m}^{max} = (\gamma-1/3)/\gamma'$. Taking into account the form of the exponent $C$ worth recalling is Lindemann – Gilvary law (Gilvary, 1966) $dT_m/dP = T_m[2(\gamma-1/2)/B]$, what indicates the pressure dependence of the power exponent in the SG-type eq. (21). Schlosser et al. eq. (13) and Kumari-Dass eq.(19) and can be extended to the negative pressures domain when introducing the reference related to the absolute stability limit in the negative pressures domain: $P \to \Delta P = P - P_{Sp}$, $V \to \Delta V = V - V_{Sp}$, $\rho \to \rho - \rho_{Sp}$.

## 3. The analysis of experimental data

When considering the parameterization of $T_g(P)$ or $T_m(P)$ experimental data, some basic problems emerges:
  (i)   Does the selected equation is proper for portraying the given set of data ?
  (ii)  What is the pressure range of applicability of the description?
  (iii) Is it possible to estimate optimal values of parameters, avoiding the uncertainty associated with the number of parameter and the nonlinear fitting ?

To address these questions in refs. (Drozd-Rzoska and Rzoska, 2006; Drozd-Rzoska et al., 2007, Drozd-Rzoska, 2005) the preliminary derivative-based and distortions-sensitive analysis of $T_m(P)$ and $T_g(P)$ experimental data was proposed: $T_g(P) \Rightarrow [d(\ln T_{g,m})/dP]^{-1}$. For SG/AA or DR equations ((6), (7), (12)) one obtains the linear behaviour of transformed experimental data (Drozd-Rzoska et al., 2007, Drozd-Rzoska, 2005):

$$\left(d\ln T_{g,m}/dP\right)^{-1} = ba + bP \quad \text{and} \quad \left(d\ln T_{g,m}/dP\right)^{-1} = b\pi + bP \tag{22}$$

It is visible that the description via DR and SG/AA relations overlaps and both can be extended into the negative pressures domain. However, such possibility for the AA and SG relation may be casual since it does not takes place for Rein&Demus and Kechin eq. (9), for Kumari&Dass eq. (19) or for pressure counterparts of the VFT relation (eqs. (2) and (3)).

Regarding the 'general' DR eq. (11), the following transformation of experimental data was proposed to test the domain of its validity (Drozd-Rzoska et al., 2008; Drozd-Rzoska et al., 2007):

$$\left[d(\ln T_m)/dP + c^{-1}\right]^{-1} = A + BP \tag{23}$$

For the optimal selection of the damping pressure coefficient $c$ one obtains the linear behaviour of transformed experimental data and the linear regression fit yields optimal values of $\pi$, $b$ and $c$ coefficients. Subsequently, they can be substituted to eq. (11), avoiding the nonlinear fitting.

Concluding, equations (22) and (23) define the way of the preliminary transformation and analysis of experimental $T_{g,m}(P)$ via the plot $d\ln T_{g,m}/dP$ vs. $P$, which indicates the domain of the domain of validity of the given description and optimal values of parameters. The derivative-based and distortions-sensitive preliminary analysis can reveal even 'weakly emergent' hallmarks of approaching $dT_{g,m}/dP > 0 \Leftrightarrow dT_{g,m}/dP < 0$ crossover, hardly 'eye-detectable'. Below, practical applications of above reasoning are discussed. First, they are focused on melting of germanium ($dT_m/dP < 0$) (Porowski et al., 2015, Vaidya et al., 1969) and subsequently for the 'soft" material, P4MP1 polymer, with $T_m(P)$ maximum (Höhne, 1999; Höhne et al., 2000). It is worth stressing that for the vast majority of systems tested so far $dT_m/dP > 0$ (Kechin, 1995; Kechin, 2001, Skripov and Faizulin, 2006) and there is much lesser number of systems where $dT_m/dP < 0$ (see Table I). Figure 2 presents such data for germanium, which can be well portrayed by DR eq. (11), with parameters obtained from the pre-analysis of experimental data via eq. (23), as shown in the inset. Notable, is the possible maximum of $T_m(P)$ curve hidden in the negative pressures domain at $P_{\max} \approx -0.32\,GPa$.

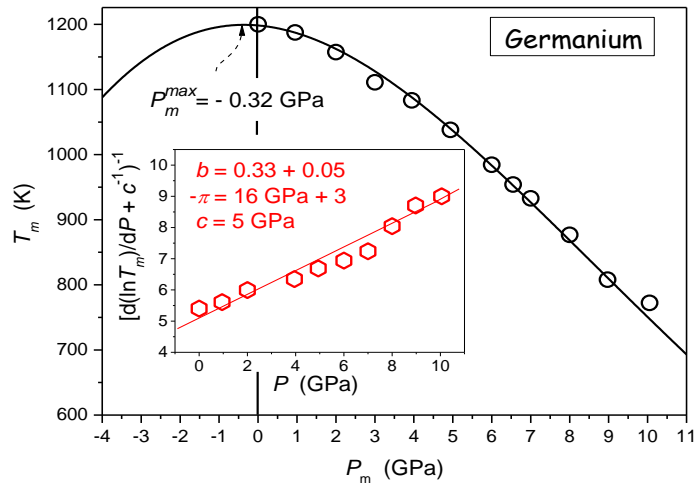

**Fig. 2** Pressure dependence of melting temperature of germanium (based on data from ref. (Vaidya, 1969; Porowski, 2015). Experimental data are portrayed by DR eq. (11),

with the support of the preliminary derivative-based analysis (eq. (23)) yielding also optimal values of parameters: this is shown in the inset.

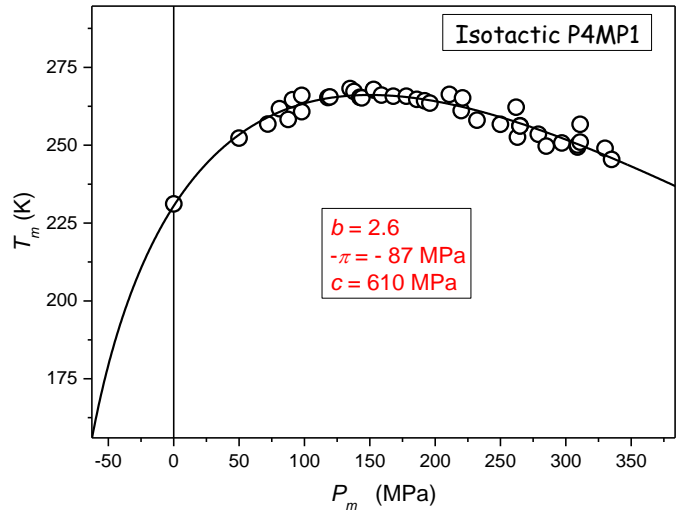

**Fig. 3** The evolution of melting temperature in poly(4-methyl-pentene-1): isotactic P4MP1 polymer: based on data from ref. (Höhne, 1999, Höhne et al., 2000) The results from eq. (11), with parameters derived due to the preliminary analysis of data via eq. (23).

Figure 3 presents the unique 'soft matter system' where the crossover $dT_m/dP > 0 \Leftrightarrow dT_m/dP < 0$ takes place at relatively low pressures: $P_{max} \approx 150 MPa$. Recalling the Kumari-Dass model (Kumari and Dass, 1988; Dass, 1985) such small value of $P_{max}$ may result from the strong pressure dependence of the Grüneissen parameter.

One can expect that different types of $T_m(P)$ dependences should be paralleled by $T_g(P)$ behaviour, taking into account the form of GFA factor. Unfortunately, the number of experimental data for $T_g(P)$ is very limited.

Fig. 4 shows the compilation of $T_g(P)$ and $T_m(P)$ experimental data available for selenium. It is notable that a single DR eq. (11) curve can describe the whole set of $T_m(P)$ data, without a hallmark of passing a liquid I – liquid II (L-L) transition (Imre and Rzoska, . This issue is worth stressing because often $dT_m/dP$ discontinuity is reported when passing the L-L transition (Imre and Rzoska, 2010). The value of $T_g/T_m$ changes $T_m/T_g(P=0.1 MPa) \approx 2/3 \rightarrow T_m/T_g(P=P_{max}) \approx 1/2$ (Drozd-Rzoska et al., 2007; Drozd-Rzoska et al., 2008). When entering the negative pressures domain the GFA factor $T_g/T_m \rightarrow 1$, i.e. the system becomes extremely good glass former.

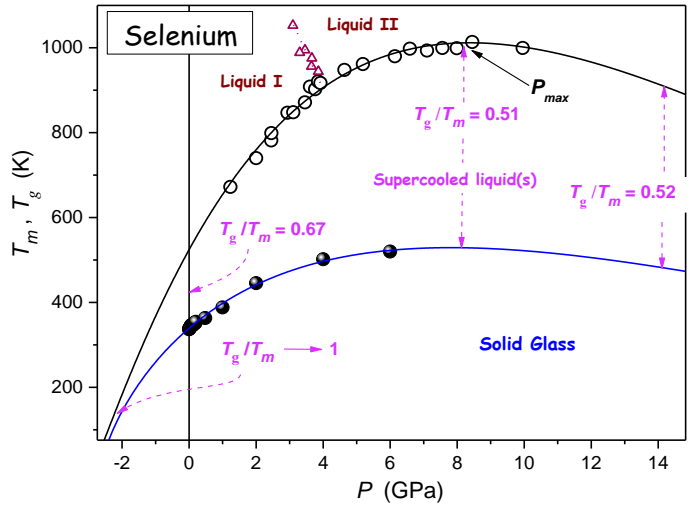

**Fig. 4** The pressure evolution of melting and glass temperature for selenium. The change of $T_g/T_m$ value is indicated. Solid curves are described by DR eq. (11): parameters were derived from the preliminary analysis based on eq. (23). Experimental data were taken from refs. (Deaton and Blum, 1965; Katayama et al., 2000; Ford et al., 1988; Tanaka, 1984; Caprion and Schober, 2002).

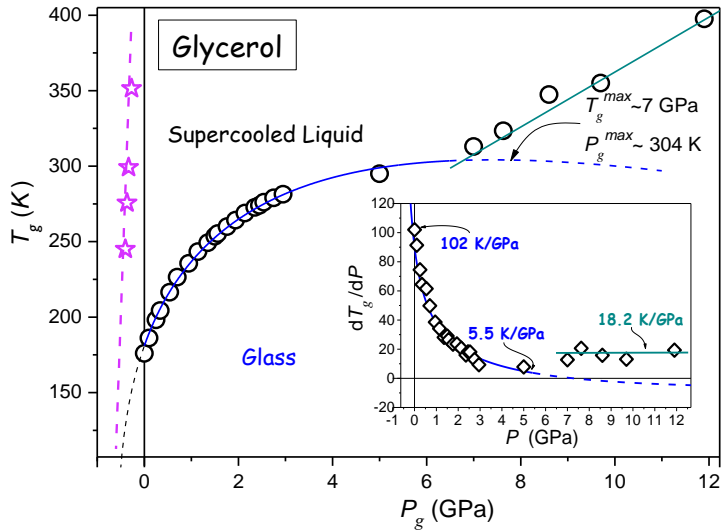

**Fig. 5** The pressure evolution of the glass temperature for glycerol. The solid blue curve, with 'dotted' and 'dashed' parts is related to DR eq. (11) and the preliminary analysis via eq. (23). Experimental data are from author's measurements [60] and from refs. (Drozd-Rzoska, 2005; Drozd-Rzoska et al., 2007, Cook, et al. 1994; Pronin et al. 2010). The dashed line and stars (in magenta) in the negative pressures domain denotes the possible absolute stability limit location: this was determined from the analysis of $\tau(P)$ experimental data via eq. (5). The inset shows the pressure evolution of $dT_g/dP$ coefficient.

Glycerol belongs to the group of the most 'classical' glass forming ultraviscous liquids (Berthier and Ediger, 2016; Rzoska et al., 2010; Donth, 2000; Rzoska et al. 2010, Rzoska and Mazur, 2007) Fig. 5 shows the compilation of data from the authors' broad band dielectric spectroscopy pressure studies and the analysis of the primary relaxation time $\tau(T,P)$ via eq. (5) supplemented by earlier $T_g(P)$ estimations (Drozd-Rzoska, 2005; Drozd-Rzoska et al., 2007). Notable is the emergence of two types of $T_g(P)$ evolution. The first one leads to the maximum of $T_g(P)$ curve at $P_g^{\max} \approx 7 GPa$ and it is followed by a hypothetical reversal vitrification associated with $dT_g/dP < 0$. However, prior to reaching the maximum, at $P \approx 6.5 GPa$ the 'cross-over' to the another form of $T_g(P)$ evolution, described by $dT_g/dP > 0$ takes place. The dashed curve shows the extrapolation of the solid blue curve, with the indication of a hypothetical 'hidden' maximum of $T_g(P)$ curve. The inset in Fig. 2 shows changes of $dT_g/dP$ coefficient on rising pressure, additionally distinguishing two different types of $T_g(P)$ evolution.

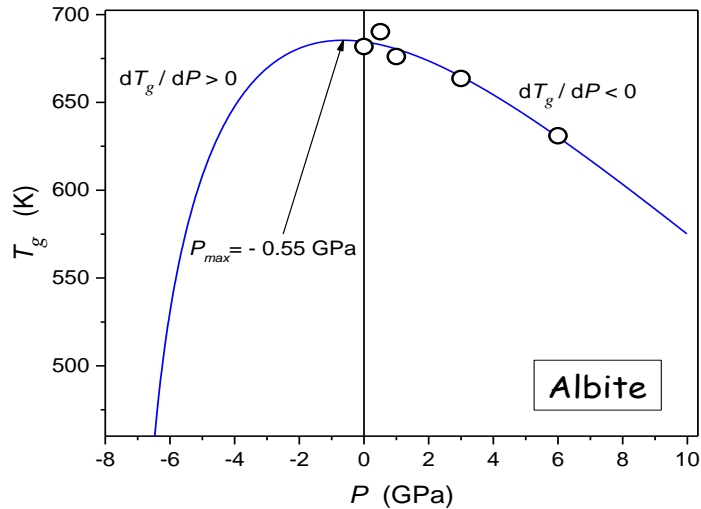

**Fig. 6** The pressure evolution of the glass temperature in albite ($NaAlSi_3O_8$), the component of magmatic, metamorphic rocks. The plot bases on experimental data from ref. (Bagdassarov, 2004). The solid curve is related to eq. (11).

Generally, the experimental evidence of glass formers characterized by $dT_g/dP < 0$ is very limited (see Table I). Such behavior seems to be characteristic for some strongly bonded systems. Fig. 7 shows results of such studies for albite, geophysically important material, which can be well portrayed by eq. (11), revealing the maximum of $T_g(P)$ curve 'hidden' in negative pressures domain.

## 4. Universal aspects of the pressure evolution of the glass temperature

The above discussion indicated the possible common phenomenological description of $T_g(P)$ evolution in glass formers described by $dT_g(P)/dP > 0$ and/or $dT_g(P)/dP < 0$. The question arises of the more microscopic insight. In ref. (Voigtmann, 2006a) analysed the vitrification within frames of the square-well (SW) model associated with the relatively simple potential: $U(r) = \infty$ for distances $r < d$ supplemented with an SW attraction within the range $\delta$, $U(r) = -U_0$ for $d < r < d(1+\delta)$ and $U(r) = 0$ beyond was analyzed. The SW approach proved its superior ability for describing colloidal glass formers, which can be thus considered as a kind of archetypical experimental glass forming model system. In ref. (Voigtmann, 2006a) the possibility of the common description of glass forming molecular liquids and colloids was shown, using the plot $\log_{10} P_g^*$ and $\log_{10} T_g^*$, where the 'natural units", i.e. model normalized glass pressure and temperature were used: $T_g^* = T_g/T_g^{model}$ and $P^* = P_g/P_g^{model}$. In ref. (Voigtmann, 2006b) the similar plot was tested for the model fluid associated with the Lennard – Jones (LJ) $V_{LJ} = 4 \in \left[ (r/\sigma)^{-12} - (r/\sigma)^{-6} \right]$ potential analyzed within the mode-coupling theory (MCT) approximation. In ref. (Voigtmann, 2005) $T_g(P)$ experimental data for glycerol, dibutyl phthalate, o-terphenyl and epoxy resin EPON 828 were analyzed ($dT_g/dP > 0$). In ref. [64] only glycerol was discussed, for the clarity of reasoning. This report also focuses on glycerol, but for the notably enhanced range of pressures, basing on data from Fig. 5. This is supplemented by experimental data for albite, where $dT_g/dP < 0$ (Fig. 6). In ref. (Voigtmann, 2006a) the SW model units were used for scaling, namely $T_g^{model} = T_g^{SW} = U_0/k_B = 826K$ and $P_g^{model} = P_g^{SW} = U_0/d^3 = 3.09 GPa$ and in ref. (Voigtmann, 2006b) the LJ model units, i.e. $T_g^{LJ} = k_B/\in = 500K$ and $P_g^{LJ} = \in/\sigma^3 = 2.5 GPa$ : numbers are given for glycerol. In ref. (Voigtmann, 2006b) the partial agreement between predictions of SW and LJ model was obtained after *ad hoc* shifting $T^* \to 1.5 T^*$. It is notable that so far experiments in colloids are carried out under atmospheric pressure and obtained phase diagrams are presented using the volume fraction ($\phi$) - interaction strength or temperature axes. Such data were model-mapped into the pressure – temperature plane in ref. (Voigtmann, 2006a). Fig. 7 recalls results of refs. (Voigtmann, 2005; Voigtmann, 2005) for: (**i**) the colloid with the addition of polymer increasing attraction and causing the 're-entrant' vitrification (Pham et al., 2002) (**ii**) glycerol ($dT_g/dP > 0$) for experimental data taken from Fig 5, (**iii**) albite for which $dT_g/dP < 0$ (Fig. 6) and (**iv**) the SW model predictions for $\delta = 0.04$ and $\delta = 0.12$ values of the key parameter, (**v**) the model using LJ potential with and without the attraction. This is supplemented by results of fitting via DR eq. (11) for glycerol and albite. One of key findings of refs. (Voigtmann 2006a and 2006b) was the 'generic steep' anomaly with exactly defined singularity, the same for any molecular glass former: $T_g^* \to 0.23$ for SW model units and $T_g^*(anomaly) \to 0.334$ for the LJ model. These led to the conclusion that there are three general regimes of glass formation resulted from $T_g(P)$ evolution (Voigtmann 2005 and 2006):

**Regime I** - for $T_g^* > 1$ : glass formers approach the hard-sphere limit. Following ref. (Voigtmann, 2006) in this domain: $T_g \propto P_g^{4/5}$ .

**Regime II** - for $1 > T_g^* > 0.23 (or\_0.334)$: there is a universal 'generic steep' anomaly and this regime is characteristic for molecular glass formers.

**Regime III** " for $T_g^* \to 0$ the low density and weak interactions domain occurs. It is available for colloidal glass formers and does not accessible for molecular ones.

In refs. (Voigtmann 2006a and 2006b) glass forming systems for which $dT_g/dP < 0$ were not discussed.

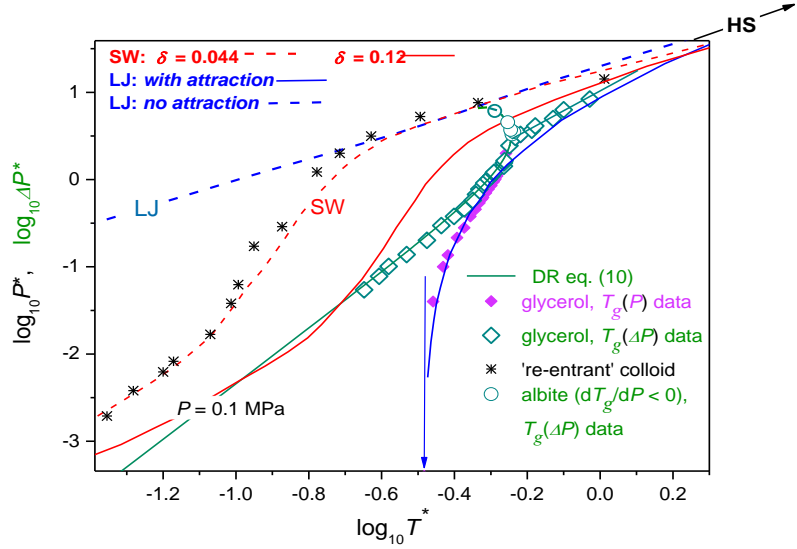

**Fig. 7** The pressure dependence of the glass temperature, summarizing the model discussion (Voirtmann 2005 and 2006): SW is for the square-well potential model, LJ – the Lennard-Jones potential model and HS is for the hard spheres model. For details see the text of the given paragraph and refs. (Voigtmann, 2006a). Experimental data for glycerol are taken from Fig. 6: they are present in the 'natural scaled" units. Data for albite are from Fig. 7. Note that for open green diamonds (glycerol) and open circles (albite) the reference pressure was takes into account: $P \to \Delta P = P + \pi$. Data for the polymer mediated colloid are from refs. (Voigtmann, 2006a; Pham, 2002). For details see comments in the given paragraph. Note the disappearance of the 'generic steep' anomaly (indicated by the vertical arrow) and the ability for describing arbitrary glass former.

One of the most striking features of refs. (Voigtmann 2006a and 2006b) is the 'generic steep' anomaly, presumably occurring only for molecular glass formers. However, this unique phenomenon has few surprising features. First, it is very strong and associated with exactly the same 'singular' value of $T_g^* \approx 0.23$ for arbitrary molecular glass former. Well above the singularity experimental data for all molecular glass formers overlaps. Second, the 'generic' anomaly appears in the log – log scale but no hallmarks of such behavior appears in the linear scale for any 'native' $T_g(P)$ data (Donth, 2000; Johari and Whalley, 1972; Drozd-Rzoska et al. 2007, Drozd-Rzoska et al. 2008; Floudas et al., 2011; Rzoska and Mazur, 2007; Andersson and Andersson, 1998; Roland, Hensel-Bielowka, et al., 2005). Third, although real high pressure results for colloidal glass formers are still not available, one can easily show that such data also will follow the same 'generic steep anomaly' pattern, in disagreement with 're-calculated' data shown in Fig. 7 (stars).

Following all these, one can conclude that the 'generic steep" anomaly is the consequence of $P \to 0$ (i.e. in practice $P \to 0.1 MPa$) within the plot applying the log-log scale. This is not a real physical phenomenon. Any fluid can be smoothly cross-overed from the hydrostatic pressures region ($P > 0$) to the isotropically stretched, negative pressures domain ($P > 0$) (Imre et al., 2002). Experimental evidences clearly show the lack of any hallmarks of passing $P = 0$, also for supercooled molecular glass formers (Imre et al., 2002; Sciortino et al., 1995; Angell and Quing, 1989). The natural termination of the liquid state is the absolute stability limit spinodal in negative pressures domain, where any liquid 'breaks' and the homogeneous cavitation occurs. Taking this as the reference one should consider the 'universal plot' based on the scale $\log_{10}\Delta P^* = \log_{10}\left[(P+\pi)/P_g^{model}\right]$ vs. $\log_{10}T_g^*$ instead of $\log_{10}P^*$ vs. $\log_{10}T_g^*$ plot. Consequently, the "generic steep" anomaly disappears and $T_g(P)$ experimental data for molecular glass formers can be mapped also to the low density ($T^* \to 0$) domain. When linking such analysis with eq. (11) one also obtains the possibility of describing systems characterized by $dT_g/dP < 0$, as shown for the extrapolated behavior for glycerol and for albite in Fig. 7. Fig. 7 also shows that the re-entrant glass forming colloids mapped from experimental studies under atmospheric pressure to the *P-T* plane are related to the case $dT_g/dP < 0$.

For glycerol, for very high pressures, the behavior described by $T_g \propto P_g^{4/5}$ emerges and the evolution approaches the hard sphere limit pattern (Voigtmann, 2006a). One of arguments for the generic importance of the 'steepness' anomaly in refs. (Voigtmann 2006a and 2006b) was the possibility of it reproduction by the model-fluid with LJ potential containing properly adjusted attraction term. However, for the analysis of $T_g^*(P_g^*)$ in such model-fluid the absolute stability limit have to be taken into account: after the transformation $P \to \Delta P$ the 'generic steep anomaly' disappears also for the LJ model fluid.

Concluding, the plot $\log_{10}\Delta P_g^*$ vs. $\log_{10}T_g^*$ offers a nice frame for the 'universal' presentation and comparison $T_g(P)$ experimental and model based data. The cross over from $dT_g/dP > 0 \to dT_g/dP < 0$ seems to be associated with $T_g^* \to 0.6$ and $T_g^* \to 3.55$ in such plot. This is the key feature of the intermediate **regime II**. There are no unique 'generic' steep anomalies. Finally, worth indicating is the general difference between $P_g^*$ vs. $T_g^*$ data taken from concentrational experiment under atmospheric pressure (1) and from the real high pressure experiment (2) for colloidal glass formers. The case (1) for re-entrant colloidal glass former can be linked to the group of systems where $dT_g/dP < 0$. The characterization of the solvent is constant but the number of colloidal particles and distances between them can change when 'decreasing pressure" ($\phi \to 0$). For such system the problem of the absolute stability limit is absent: it is naturally related to $P_g^* = 0$ and the negative pressures domain does not exist. For the case (2), compressing changes notably not only not only distances between colloidal particles but also properties of the solvent. Changes of density of the solvent (typically ~ 30 % for $P \approx 1GPa$) are associated with very strong changes in dynamics, particularly near the glass temperature. In this case 'rarefication' associated with the isotropic stretching and entering pressures domain can yield even stronger changes for the solvent. Stretching is terminated by the absolute stability limit spinodal in negative pressures domain. All these show that for the case (1) properties of the colloidal glass former are dominated almost exclusively by colloidal particles. In the case (2) at least equally important is the impact of the solvent.

Fig. 7 indicates the clear link between molecular and colloidal glass formers: they follow the same patter the plot $\log_{10}\Delta P_g^*$ vs. $\log_{10}T_g^*$. Model fluids based on SW and LJ potentials offer the nice frame for getting the fundamental insight into experimental data within such presentation.

## 5. Concluding remarks

This report presents proposals of few equations for describing the pressure evolution of the glass temperature beyond the dominated SG/AA pattern. They make the description of glass forming systems where both $dT_g/dP > 0$ and $dT_g/dP < 0$ possible. The ways of portrayal were extended also for the evolution of $T_g(V,\rho)$ and $P_g(V,\rho)$. The basic relevance of including into the analysis negative pressures and the preliminary derivative-based and distortions – sensitive analysis has been shown. From results presented the possible general pattern for $T_g(P)$ evolution for glass forming systems ranging from low molecular weight liquids, resins, polymer melt, liquid crystals to colloidal fluids emerges.

In the low density region the extended SG-type equation can describe experimental data. On increasing pressures, for intermediate densities, the gradual inclusion of the 'damping term' can lead to the reversal (re-entrant, $dT_g/dP < 0$) vitrification. However, for strongly compressed and high density systems the crossover to the second, HS-type, dependence $T_g(P) \to P_g^{4/5}$ takes place. The cross over to this second type of vitrification can occur before reaching the maximum of $T_g(P)$ as for glycerol or well beyond the maximum. poiFor the model-normalized 'universal' plot $\log_{10} \Delta P_g^*$ vs. $\log_{10} T_g^*$ such general characterization is manifested as the less or more marked *S-shape*. It is notable that this picture may be valid both for molecular and colloidal glass formers, although for the latter real high pressure experiments are still required. For the dominated group of systems where $dT_{g,m}/dP > 0$ most often the SG/AA-type ($T_{g,m}(P)$), Kraut-Kennedy – type ($T_{g,m}(V,\rho)$ or Murngham – type ($P_{g,m}(V,\rho)$)) dependences are used. The discussion for the latter (Poirier, 2000; Skripov and Faizulin, 2006) indicate that notable distortions appears for $\Delta V/V_0 \to 1/2$. Taking into account the compressibility of typical molecular liquids such domain starts for $P \sim 1.5 GPa$. In the opinion of the authors, equally important can be the distance of the reference point from the possible maximum of $T_g(P)$, even if it is 'hidden' by a phase transition or crossover to another form of vitrification.

Finally, we would like to stress the significance of the above discussion for the glass transition physics, material engineering and geophysical and planetary studies.

## 6 Conflict of Interest

The authors declare that the research was conducted in the absence of any commercial or financial relationships that could be construed as a potential conflict of interest.

## 7. Funding

This report was prepared due to the support of the National Centre for Science (Narodowe Centrum Nauki (NCN), Poland) grant ref. UMO-2016/21/B/ST3/02203.

**Figures Captions**

**Figure 1** The temperature dependence ($P = 0.1$ MPa) of the proper volume $V = 1/\rho$, $\rho$ denotes density, for polyvinyl acetate (PVA) in the ultraviscous and solid amorphous phases. Dashed lines show extrapolations of the experimental behavior remote from the 'stretched' glass transition domain $T_g$. The apparent discontinuity of the volume can be estimated as $\Delta V' = 0.0021 cm^3 g^{-1}$ and $\Delta V'' = 0.0030 cm^3 g^{-1}$ (double arrows in the plot). The inset, based on data from ref. (McKinney, 1974; Tropin, 2012; Roland and Casalini, 2003), is for the excess of the specific heat $\Delta c_p(T) = c_P^{melt}(T) - c_p^{solid}(T)$, over the behavior in the solid stated remote from $T_g$ $c_p^{solid}(T) = a + bT$ described the behavior well below. The resulting discontinuity $\Delta c_p(T)/R = 0.23$. Data in Fig. 1 are for 10 K/min. cooling / heating rate.

**Figure 2** Pressure dependence of melting temperature of germanium (based on data from ref. (Vaidya, 1969; Porowski, 2015). Experimental data are portrayed by DR eq. (11), with the support of the preliminary derivative-based analysis (eq. (23)) yielding also optimal values of parameters: this is shown in the inset.

**Figure 3** The evolution of melting temperature in poly(4-methyl-pentene-1): isotactic P4MP1 polymer: based on data from ref. (Höhne, 1999, Höhne et al., 2000) The results from eq. (11), with parameters derived due to the preliminary analysis of data via eq. (23).

**Figure 4** The pressure evolution of melting and glass temperature for selenium. The change of $T_g/T_m$ value is indicated. Solid curves are described by DR eq. (11): parameters were derived from the preliminary analysis based on eq. (23). Experimental data were taken from refs. (Deaton and Blum, 1965; Katayama et al., 2000; Ford et al., 1988; Tanaka, 1984; Caprion and Schober, 2002).

**Figure 5**   The pressure evolution of the glass temperature for glycerol. The solid blue curve, with 'dotted' and 'dashed' parts is related to DR eq. (11) and the preliminary analysis via eq. (23). Experimental data are from author's measurements [60] and from refs. (Drozd-Rzoska, 2005; Drozd-Rzoska et al., 2007, Cook, et al. 1994; Pronin et al. 2010). The dashed line and stars (in magenta) in the negative pressures domain denotes the possible absolute stability limit location: this was determined from the analysis of $\tau(P)$ experimental data via eq. (5). The inset shows the pressure evolution of $dT_g/dP$ coefficient.

**Figure 6**   The pressure evolution of the glass temperature in albite ( $NaAlSi_3O_8$ ), the component of magmatic, metamorphic rocks. The plot bases on experimental data from ref. (Bagdassarov, 2004). The solid curve is related to eq. (11).

**Figure 7**   The pressure dependence of the glass temperature, summarizing the model discussion (Voirtmann 2005 and 2006): SW is for the square-well potential model, LJ – the Lennard-Jones potential model and HS is for the hard spheres model. For details see the text of the given paragraph and refs. (Voigtmann, 2006a). Experimental data for glycerol are taken from Fig. 6: they are present in the 'natural scaled" units. Data for albite are from Fig. 7. Note that for open green diamonds (glycerol) and open circles (albite) the reference pressure was takes into account: $P \to \Delta P = P + \pi$. Data for the polymer mediated colloid are from refs. (Voigtmann, 2006a; Pham, 2002). For details see comments in the given paragraph. Note the disappearance of the 'generic steep' anomaly (indicated by the vertical arrow) and the ability for describing arbitrary glass former.

**Table I  caption**

Examples of systems in which the application of pressure decreases the glass temperature ( $dT_g/dP < 0$ ) [31-38]. For the dominant group of glass formers (molecular liquids, polymers, ..): $dT_g/dP > 0$ (Donth, 2000; Floudas et al. 2011, Roland et al., 2005).